# Detecting Centralized Architecture-Based Botnets using Travelling Salesperson Non-Deterministic Polynomial-Hard problem, TSP-NP Technique

**Victor R. Kebande**
Computer Science & MT
Malmö University
Malmö, Sweden
victor.kebande@mau.se

**Richard Adeyemi Ikuesan, Abdullah Al-Ghushami**
Cybersecurity and Networking Department,
School of Information Technology,
Community College of Qatar, Doha, Qatar
{richard.ikuesan, abdullah.alghushami}@ccq.edu.qa

**Nickson M. Karie**
*Security Research Institute*
*School of Science-Edith Cowan University*
Joondalup, Australia
n.karie@ecu.edu.au

**H. S. Venter**
Department of Computer Science,
Faculty of Engineering, Built Environment and
Information Technology,
University of Pretoria, Pretoria, South Africa
hventer@cs.up.ac.za

*Abstract*— The threats posed by botnets in the cyberspace continue to grow each day and it has become very hard to detect or infiltrate the cynicism of bots. This, is owing to the fact, that, the botnet developers each day, keep changing the propagation and attack techniques. Currently, most of these attacks have been centered on stealing computing energy, theft of personal information and Distributed Denial of Service (DDoS) attacks. In this paper, the authors propose a novel technique that uses the Non-Deterministic Polynomial-Time Hardness (NP-Hard Problem) based on the Traveling Salesperson Person (TSP) that depicts that a given bot, $b_j$, is able to visit each host on a network environment, *NE*, and then it returns to the botmaster, in form of instruction(command), through optimal minimization of the hosts that are (may) be attacked. Given that $b_j$ represents a piece of malicious code and TSP-NP Hard Problem, which forms part of combinatorial optimization, the authors present this as an effective approach for the detection of the botnet. It is worth noting that the concentration of this study is basically on the centralized botnet architecture. This holistic approach shows that botnet detection accuracy can be increased with a degree of certainty and potentially decrease the chances of false positives. Nevertheless, a discussion on the possible applicability and implementation has also been given in this paper.

## I. INTRODUCTION

With the increasing sophistication of modern technologies,the world is, on the other hand, experiencing a significant increase in botnet attacks such as Distributed Denial of Service (DDoS), malware dissemination, and phishing attacks among other types of botnet-based attacks. Additionally, the ability to want more computing power has made people find ways of controlling IT resources which in reality do not even belong to them. It is on these grounds that it has enabled adversaries to utilise botnets as threat agents for malicious gains.

Cyber-adversaries have the ability to give botnets such a high computing power to use for whatever reason they please-ranging from financial gains, service disruption, DDoS and sensitive information theft. In most cases, anyone who creates a botnet has the capability of controlling and managing the infected computers on the network remotely over the Command and Control (C&C) infrastructure, with virtually no risk of being detected in some instances. This is because the owners of the infected computers (zombies) are often unaware that they are infected and that someone somewhere is controlling them.

The detection of botnets can be very hard. It is for this reason that we propose in this paper a technique that uses the Non-Deterministic Polynomial-Time Hardness (NP-Hard Problem) based on the Traveling Salesperson Person (TSP), that shows diverse traversal techniques, in order to detect the presence of a botnet through optimal minimization of the host endpoints.

As for the remaining part of this paper, Section 2 covers the background while section 3 handles the related works. Thereafter, Section 4 presents an overview of the proposed approach followed by a discussion in Section 5. Finally, the paper concludes in Section 6 and makes mention of future work.

## II. BACKGROUND

This section presents a background study of the following areas: Botnets, Centralized Botnet Architecture and the TSP-NP-Hard Problem.

## A. Botnets

A botnet can simply be understood as a robot network. In a botnet, compromised computers or IT infrastructure are put under the control of a malicious actor. Each device in a botnet is referred to as a bot and can run autonomously and automatically. However, bots can also be addressed as "zombie computers" due to their ability to operate under remote direction from a malicious actor without their owners' knowledge once a victims' computer is compromised.

Research by Liu et al.,[1] shows that botnets have become prevalent mostly in wired as well as wireless networks and also the cloud. Also, internet technology plays a big role in the spread and control of exiting botnets. Besides, the Internet has also made it very hard to detect botnets due to their evolving and complex techniques used by botnet controllers to avoid detection.

Compared to other existing malware, the authors in [2] argue that botnets are emerging as the most serious threat against cybersecurity, as they provide a distributed platform for several illegal activities. This includes but not limited to launching DDoS attacks against critical targets, malware dissemination, phishing as well as click fraud. Other research in botnet detection strategies and propagation has been mentioned in [4-7].

In this paper, the authors argue that the threats posed by botnets in cyberspace are increasing each day and it has become very hard to detect or infiltrate bots given that the botnet developers keep changing propagation and attack techniques. It is for this reason that this paper concentrate on proposing an effective approach for botnet detection based on TSP-NP-Hard Problem that is based on centralized botnet architecture. The next section will briefly explain the centralized botnet architecture.

## B. Centralized Botnet Architecture

For a very long time, a bigger percentage of created botnets have had a common centralized architecture [3]. In essence, centralized botnet architecture means that the bots in the botnet connect directly to some special hosts also called "command-and-control" (C&C) servers under the control of the botnet operator.

The primary work of the C&C server is to receive commands and signals from the operator named the botmaster, and then propagate them to other bots in the same network automatically [4]. Figure 1 shows how a centralized botnet architecture would look like.

Figure 1, infer that the C&C server is responsible for monitoring the status of all connected bots, issues commands originating from the botmaster, in order to collect data and constantly wait to connect new bots as well as registering them in its database for further attacks.

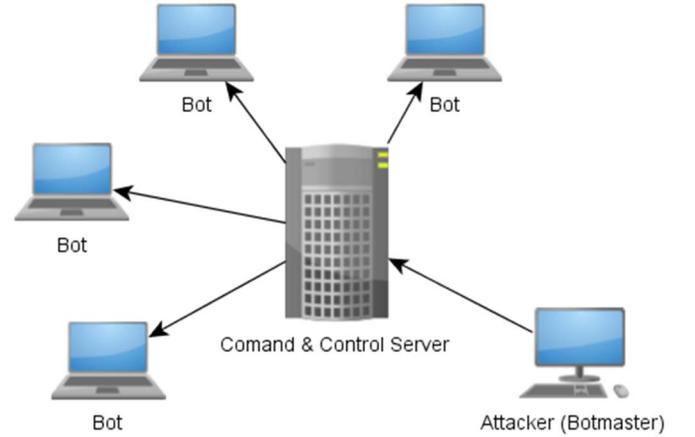

**Fig 1: Centralized botnet architecture**

Based on the centralized botnet architecture, the authors propose in this paper a technique that uses the Non-Deterministic Polynomial-Time Hardness (NP-Hard Problem) based on the Traveling Salesperson Person (TSP) to show that a given Bot, $b_j$, is able to visit each host on a network environment, NE, and then returns to the botmaster in form of instruction through optimal minimization of the hosts that are attacked. The TSP-NP-Hard Problem is briefly discussed in the subsection to follow.

## C. TSP-NP-Hard Problem

A Traveling Salesperson Problem (TSP) is represented given $\frac{(n-2)}{n/2}$ integers based on the $n$ pair of traversals, where all cities can be visited and back to the origin [8]. TSP problems were first realized in 1930 by Viennese Mathematician Karl Menger. The TSP tries to find the shortest path through joining a finite set of points with the distance between them given. One point could easily be determined based on a particular ordering $i_1, i_2 \ldots \ldots \ldots, i_{1n+1}$ and the length of the distance can be given by $C_{i_1 i_2}, C_{i_2 i_2} \ldots \ldots \ldots, C_{1n+1, i_2+1}$ where the rules of the origin depend on the number of permutations for the cities to be traversed [9],[10].

The authors consider the following TSP-NP Hard problem definitions:

***Definition 1***: *If for the set of vertices a, b, c $\in V$, it is true that t (a, c) $\leq$ t(a, b) + t(b, c) where t is the cost function, we say that t satisfies the triangle inequality* [11].

A Minimum Spanning Tree (MST) represents the weight of which is a lower bound on the cost of an optimal traveling salesman tour. Using this MST, one would create a tour of the cost of which is at most 2 times the weight of the spanning tree.

***Definition 2***: *If an NP-complete problem can be solved in polynomial time then P = NP, else P ≠ NP* [11].

Given a problem, there can be an approximation for the solution based on the cost of the optimal solution. If the cost is not able to satisfy the given traversal (inequality) then the polynomial-time may not be enough to find the acceptable approximation to the traversals across the hosts/vertices.

***Definition 3***: *The domination number for the TSP of a heuristic A, is an integer such that for each instance I, of the TSP on n vertices A, produces a tour T, that is now worse than at least d(n) tours in I including T. This also includes the greedy approaches and the aspect of the Nearest Neighbor (NN)* [11].

The TSP also depends on the optimization of the greedy approaches and the aspects of the nearest neighbors which helps in the traversals.

### III. RELATED WORK

Research in [12] has solved the NP-complete problem of an Ant Colony Optimization (ACO), which has been able to compare the performance matrix of ACO in order to solve the Soduko Puzzle. While this research employs the TSP-NP Hard problem, no focus has been given entirely on the botnet detection strategies at the time of writing this paper. Nevertheless, research by [13] has formally presented a decision problem where if $A$ is a decision problem, then $A$ has a succinct certificate property, $A \in NP$ and $A$ is transferable to Integer Programming in polynomial time. Based on this problem, the authors have been able to deduce also that $P \subseteq NP$, then if $A \in P$, then $A \in NP$. Another research in [14] has brought insights into the design of parameterized complexity analysis of heuristics. Important to highlight is that in that research, an ACO approach for Euclidean TSP has been used and a more effective ACO algorithm is able to be obtained.

### IV. DETECTING CENTRALIZED ARCHITECTURE-BASED BOTNETS USING TRAVELLING SALESPERSON NON-DETERMINISTIC POLYNOMIAL-HARD PROBLEM-TSP-NP TECHNIQUE

This section gives the proposed approach that forms the main contribution of this research study. This approach concentrated on finding suitable techniques for detecting centralized architecture-based botnets based on the Traveling Salesperson Problem Non-Deterministic Polynomial-Hard problem-TSP-NP.

#### A. Problem Formulation

The problem has been formulated based on the following approach: Let $x$ be the number of nodes in a network and $n$ be the number of hosts that a botnet can attack in a network environment, $NE$. For each $n$, in a network, there exists $b_j$ bots that form pieces of malicious code that can compromise a number of hosts and form part of the remote-controlled entities that constitute a botnet, $Bt$. The $b_j$ are represented as entities that can attack an existing number of hosts. The author takes a single host, $h_i$ as a primary attack point that $b_j$ can attack. Next, the author refers to the $K^{th} -$ attack on $h_i$ $S.T$ $h_i, k = 1 \dots k_{n+1}$. We then define the probability of an attack on $h_i$ based on the number of bots that are dispatched and the defense mechanism of the victim's knowledge of the attack. A victim can easily suffer a botnet attack if one is able to trigger a malicious code to be executed unknowingly. An assumption is made that there is a probability, $P_i$ that any $h_i$ can be compromised and each $b_j$ can be taken to be independent. Based on this assumption, in the attacker's perspective, $P_i$ that $h_i$ can be attacked depends on the propagation techniques. The objective function on the TSP propagation techniques that are represented as follows:

$$\sum P_i(b_i)x_i \tag{1}$$

Which is the probability function $P_i$ that shows that $h_i$ can be attacked. The notations have been represented as is shown (Table 1).

**Table 1. Notations**

| Notation | Description |
|---|---|
| $P_i$ | The probability that a host may be attacked |
| $b_i$ | Possible defense mechanisms for botnet where $i \in N$ |
| $x_i$ | Returns 1 if a host is attacked by a botnet otherwise it returns 0 |
| $w_{xn+1}$ | Weight function of a network/directed/undirected graph |

#### B. TSP-NP-Hard Problem Approach in Botnet Detection

The process of TSP is represented as a complete graph:

$TSP\{G, f, t\}: G(V, E)$ where $G$ is a graph, $f$ is a function, $VxV \rightarrow Z$ and $t \in Z$.

$G$ is also a graph with vertices $V$ and edges $E$ over which a botmaster can be able to dispatch malicious codes (bots) between $V$ and $E$, which are also used as end-points and hosts respectively.

The authors take the TSP-NP hard problem approach where a botmaster can initiate bot traversals and the bots can communicate back to the botmaster. This has been shown in Figure 2. Based on the TSP end-points in Figure 2, a botmaster can traverse the shortest route through which $b_i$ can attack $h_i$ either through $TSP\ [B\{x_i\}, B\{x_{ii}\}, B\{x_{ii}\}, B\{x_{iv}\}, B\{x_{iv}\}]$.

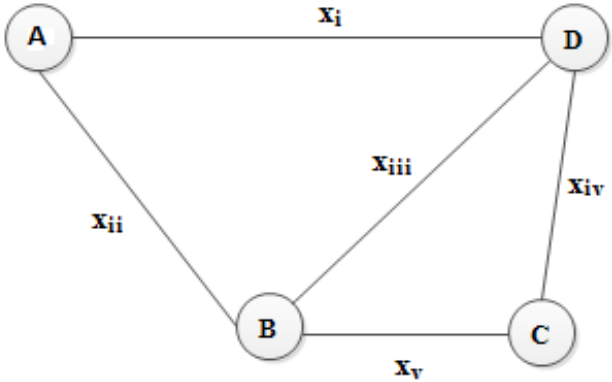

**Fig. 2: Weighted TSP end-points**

The TSP proof lies in showing that TSP belongs to NP. To check this, the botmaster can take a tour (injection) and a bot can propagate on each vertex once and then communicate back to the botmaster.

*C. Outline-TSP-Botnet-Attack-Detection Proof*

The authors make an assumption based on how the bots are able to propagate in a network. This shows that there is always $h_i$ that may be vulnerable to attacks. The proof for the TSP-Attack is represented using *Lemma 1, 2* and *3* as follows:

**Lemma 1:** *If there is no defense mechanism in $h_i$ during initial bot injection and secondary infection by $b_j$, then $b_j$ can be successful, given that $b_j \in N$ where the $b_j$ binaries are able to be executed in $h_i$ to perform a wide range of cynical activities.*

**Proof:** We denote $\{b_i, b_2, \ldots \ldots b_{j+1}\}$ as a set of bots $S.T\ b_j \in N$ that are executed at the node end-points or the hosts. Based on this, a set of bots that are delivered to extract information from $\{h_i, h_2, \ldots \ldots h_{n+1}\}$ may be executed as follows: $[< b_1 \to h_1 >, < b_2 \to h_2 \ldots \ldots \ldots \ldots, < b_{j+1} \to h_{n+1} >]$. We can also represent $\{t_i, t_2, \ldots \ldots t_{j+1}\}$ as the time it takes for a bot to travel from one host (vertice) to another and back to the botmaster.

The bots are executed as the hosts and based on TSP, there is a frequent message exchange between the botmaster and the bots and to finalize this communication, there is usually a message exchange $[< b_{j=1} \to h_{n+1} >] \to Botmaster$ in order to complete the process: $TSP: VxV \to Z\ approach$.

**Lemma 2:** *A bot $b_j$ traversing a host, $h_i$ (vertices) will propagate based on the time $\{t_i, t_2, \ldots \ldots t_{j+1}\}$ and no update from a botmaster may be missed.*

**Proof:** We consider a situation where a propagating bot fails to hit the target host. This may require the botmaster to send more bot updates. A newer version of a bot will replace the existing and the code will be executed, hence no bot update is likely to be missed, (*See Lemma 1*).

**Lemma 3:** *Once a bot is successfully executed inside a host (vertices), the bot can change its state of execution to appear to be deterministic as possible.*

**Proof:** This implementation shows that based on TSP-NP, it is indeed deterministic based on how a bot behaves after execution. This can be shown based on how $[< b_{j=1} \to h_{n+1} >]$ behaves over $\{t_i, t_2, \ldots \ldots t_{j+1}\}$ (*see lemma 1 and lemma 2*).

*D. Optimization of TSP-NP Botnet Detection using Greedy*

Based on the botnet traversal and propagation techniques [*see, Lemma 1, 2 and 3*], we access $TSP - NP - hard\ problem$ based on how the bot traverses or travels once it is dispatched using $[Botmaster - -C\&C\ server < b_{j=1} \to h_{n+1} >] \to C\&C\ server \to Botmaster$ to host and back as is shown in Figure 3.

An assumption is made that the TSP-NP hard problem is represented using $T_{Bn}$ traversals, where $Bn$ represents the botnet in a directed/weighted graph $G = (V, E)$ with weights $w(x1), w(x2), w(x3)\ and\ w(x4)$ respectively. This also commensurate with infection happening every time a bot moves from $h_n[b_{j+1}]$ hosts and when all $h_n[b_{j+1}]$ are visited, instructions can be dispatched back to the origin *botmaster* using $TSP \to C\&C \to Botmaster$. We optimize this solution using greedy approaches using the following assumptions

1. Assuming that a bot that is dispatched by a botmaster can have the worst tour, then it is imperative to have exceptions.
2. Assume that all the botnet traversals happen between the edges, then there needs to exist a Nearest Neighbor (NN) that has a cost based on the starting point which is 1 if there exists an NN or 0 otherwise.

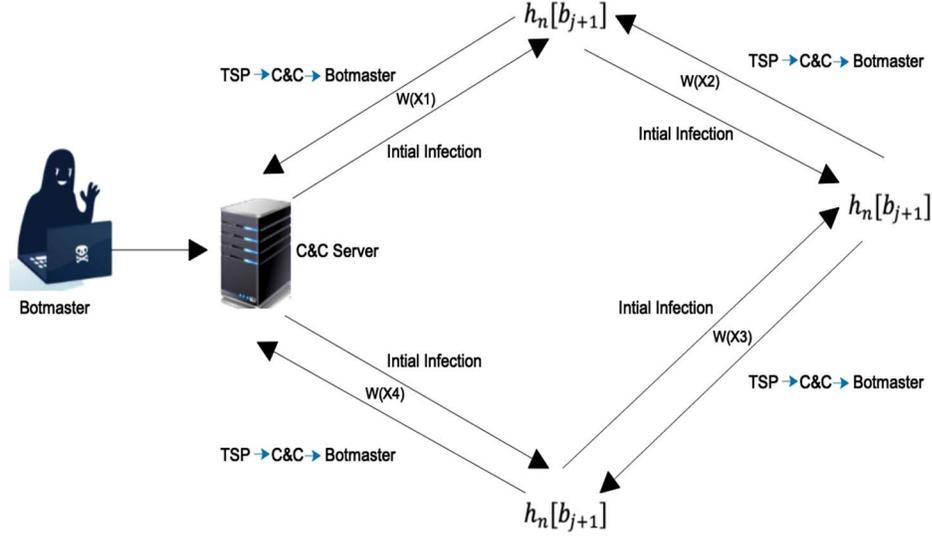

**Fig .3: Centralized TSP Botnet Traversals**

3. A consideration that the TSP-for bots prove that the NN has domination of bots on the hosts

Based on the three assumptions that are labeled [1-3] above we coin the following botnet traversal equations [2 to 6].

$$[T_{Bn}, W_{(xn+1)}] \qquad (2)$$

Which shows the bot travels and the distance covered based on TSP-NP as shown in Equation 3

$$\sum_{i=1}^{n-1}(X_{n+1} + T_{Bn}) \qquad (3)$$

which shows the exceptions in the worst tour as,

$$\sum_{k=1}^{n-1} kN - N + 1 \qquad (4)$$

then it shows the NN node that a bot is likely to visit as,

$$\sum_{k=1}^{n-1} kN > 1 \qquad (5)$$

If the process begins with a 1 if a NN exists

$$C_{NN} = C[1,2 \ldots \ldots n] = \sum_{i=1}^{n-1} iN + N + 2 \qquad (6)$$

Represents the cost to the TSP-NP hard problem that a bot may take in the NN or the nearest host/node. Having looked at the TSP-NP Hard problem approaches that can be used to aid in centralized botnet detection, in the next section a discussion of this study is given.

## V. DISCUSSION

The authors have mainly focused on two distinct issues: First, the attention of the study has focused on the assessment of the centralized botnet topology and how easily the bot organization can influence major botnet attacks. Consequently, the authors have been able to give this study a TSP-NP-hard problem approach in order to foster botnet detection. Normally in a centralized botnet architecture, the most important aspect is to know where the C&C server sits because it is the central point through which a botmaster uses to dispatch instructions to bots and it is the same point over which bots can report to the botmaster about the stolen information.

Given that the centralized architecture is organized using point/nodes that are connected [15-16], the attacker bots begin to attack nodes from a point $x$, then they are able to traverse to different nodes $\{x_1 x_2 x_3 \ldots \ldots \ldots \ldots x_{n+1}\}$, through a visit of all the nodes in the shortest path possible and then, convey the message back to the botmaster in a TSP-NP hard problem approach. This means that the attacker needs to compromise a number of nodes. Based on the TSP-NP hard problem approach, to detect the influx of bots at nodes/hosts, the cyber-defender would be required to easily pay attention to the node-traversing patterns and the node characteristics. This is owing to the fact that the TSP-NP hard problem allows a bot to visit each node/host depending on the instruction from the botmaster, then the characteristics that may include the ability for the bot to do phishing attack, click fraud and malicious attachments, etc. This could easily be detected based on how they traverse on each node. It is the authors' opinion that the proposed TSP-NP-Hard problem approach is effective enough and it also provides a holistic approach that can easily be employed in an attack and defense-based scenarios.

As part of the implementation of this study, the authors aim to simulate a network by creating an army of zombies and then use the TSP-NP and NN to identify and provide security of the network against botnets based on DDoS, Internet Relay Chat (IRC), click-fraud, port scan and identification of worm traffic from a C&C server perspective. This enables one to be able to profile an adversary based on the cynicism that is identified on the network based on the botnet attack characteristics. Forensic modeling and analysis [17]–[19] of these attack characteristics can then be carried out to ascertain the feasibility of investigating such an attack. Furthermore, this process will entail the development of a proactive forensic framework which can be leveraged to address such attack. Whilst the potential of such an attack presents a daunting challenge, the development of such a proactive approach could provide a baseline for addressing botnet-based cyber-incident. Modeling of the resulting behavioral characteristics of each contributing zombie can provide a pointer for understanding such an attack. A subsequent study will attempt to provide insight into methods and approaches that can be used to prevent zombie-net. This is essential in a typical user identification process, and forensic investigation, as false positives can be easily induced. Attempt to minimize the potential of false error rates during the investigation of such centralized attack architecture will also be explored.

## VI. CONCLUSION AND FUTURE WORK

This research study has mainly employed the TS-NP- non-deterministic polynomial Hard Problem to aid in the detection of the centralized-based botnets. The authors have presented this approach using mathematical approaches which have also summed up as a major contribution of this research. For future work, the authors aim to extend this technique to Peer-to-Peer botnet architecture with real-time attack-defense scenarios within a network.

## ACKNOWLEDGMENT

The authors would like to acknowledge the support of Community College of Qatar, Malmö university, and Edith-Cowan University for their respective contributions towards the realization of this research.

## REFERENCES

[1] J. Liu, Y. Xiao, K.Ghaboosi, H. Deng and J. Zhang, "Botnet: Classification, Attacks, Detection, Tracing, and Preventive Measures". Journal on Wireless Communications and Networking 2009: 692654. https://doi.org/10.1155/2009/692654

[2] M. Feily, A. Shahrestani & S. Ramadass, "A Survey of Botnet and Botnet Detection," 2009 Third International Conference on Emerging Security Information, Systems and Technologies, Athens, Glyfada, 2009, pp. 268-273.

[3] P. Wang, S. Sparks and C. Zou."An Advanced Hybrid Peer-to-Peer Botnet". 2007 Available at: https://www.usenix.org/legacy/event/hotbots07/tech/full_papers/wang/wang_html/ [accesed 12 Dec, 2018]

[4] V.R. Kebande, "Novel Cloud Forensic Readiness Service Model". 2018- [Accessed] at https://repository.up.ac.za/handle/2263/66140

[5] V.R. Kebande and H.S. Venter, "A cognitive approach for botnet detection using Artificial Immune System in the cloud". In *Cyber Security, Cyber Warfare and Digital Forensic (CyberSec), 2014 Third International Conference on*(pp. 52-57). 2014, IEEE.

[6] V.R. Kebande and H.S. Venter, "Obfuscating a cloud-based botnet towards digital forensic readiness. In *Iccws 2015-The Proceedings of the 10th International Conference on Cyber Warfare and Security* (p. 434), 2015.

[7] V.R. Kebande and H.S.Venter, "A cloud forensic readiness model using a Botnet as a Service". In *The International Conference on Digital Security and Forensics (DigitalSec2014)* (pp. 23-32). The Society of Digital Information and Wireless Communication, 2014.

[8] C.H.Papadimitriou,"The Euclidean travelling salesman problem is NP-complete". 1977 *Theoretical computer science*, *4*(3), 237-244.

[9] M. Held, A.J Hoffman, E.L Johnson, and P.Wolfe , "Aspects of the traveling salesman problem. *IBM journal of Research and Development*, *28*(4), 1984, 476-486.

[10] Karl Menger, Ergebnisse eines Kolloquiums 3, 11-12 (1930).

[11] The Traveling Salesman Problem. Accessed-https://www.csd.uoc.gr/~hy583/papers/ch11.pdf

[12] M. Asif and R. Baig, "Solving NP-complete problem using ACO algorithm". In *Emerging Technologies, 2009. ICET 2009. International Conference on* (pp. 13-16). IEEE.

[13] E. L. Lawler, J. K. Lenstra, A. H. G. Rinnooy Kan and D. B. Shmoys. The Traveling Salesman Problem: A Guided Tour of Combinatorial Optimization. John Wiley & Sons, 1985

[14] S. Nallaperuma, A. M. Sutton and F. Neumann, "Parameterized complexity analysis and more effective construction methods for ACO algorithms and the euclidean traveling salesperson problem". In *Evolutionary Computation (CEC), 2013 IEEE Congress on* (pp. 2045-2052). IEEE.

[15] V.R. Kebande and H.S. Venter, "Novel digital forensic readiness technique in the cloud environment" *Australian Journal of Forensic Sciences*, *50*(5), 2018, 552-591.

[16] V.R. Kebande and H.S. Venter, "On digital forensic readiness in the cloud using a distributed agent-based solution: issues and challenges". *Australian Journal of Forensic Sciences*, *50*(2), 2018, 209-238.

[17] A. R. Ikuesan and H. S. Venter, "Digital behavioral-fingerprint for user attribution in digital forensics: Are we there yet?," Digit. Investig., vol. 30, pp. 73–89, 2019.

[18] D. Ernsberger, A. R. Ikuesan, H. S. Venter, and A. Zugenmaier, "A Web-Based Mouse Dynamics Visualization Tool for User Attribution in Digital Forensic Readiness," in 9th EAI International Conference on Digital Forensics & Cyber Crime, 2017, pp. 1–13.

[19] M. Mohlala, A. R. Ikuesan, and H. S. Venter, "User attribution based on keystroke dynamics in digital forensic readiness process," in 2017 IEEE Conference on Applications, Information and Network Security, AINS 2017, 2018, vol. 2018-Janua, pp. 1–6.